\documentclass[aps,prb,amsmath,amssymb,superscriptaddress,reprint,
]{revtex4-2}
\usepackage{amsmath}

\usepackage{graphicx}
\usepackage{dcolumn}
\usepackage{bm}
\usepackage{hyperref}
\hypersetup{colorlinks=true,allcolors=blue}

\usepackage{siunitx}
\usepackage{physics}
\usepackage{color}
\usepackage{mathptmx}
\usepackage[normalem]{ulem}
\usepackage{braket}

\begin{document}

\title{Multiple quantum spin Hall states and topological current divider in Twisted Bilayer WSe$_{2}$} 

 
\author{Hao He}
\altaffiliation[These authors ]{contributed equally to this work.}

\affiliation{Hebei Provincial Key Laboratory of Photoelectric Control on Surface and Interface, School of Science, Hebei University of Science and Technology, Shijiazhuang, 050018, China}

\author{Zhao Gong}
\altaffiliation[These authors ]{contributed equally to this work.}

\affiliation{Hebei Provincial Key Laboratory of Photoelectric Control on Surface and Interface, School of Science, Hebei University of Science and Technology, Shijiazhuang, 050018, China}

\author{Shuai Li}

\affiliation{Hebei Provincial Key Laboratory of Photoelectric Control on Surface and Interface, School of Science, Hebei University of Science and Technology, Shijiazhuang, 050018, China}

\author{Jian-Jun Liu}
\affiliation{College of Physics, Hebei Normal University, Shijiazhuang 050024, China}
\affiliation{Department of Physics, Shijiazhuang University, Shijiazhuang 050035, China}

\author{Hui-Ying Mu}
\email[Correspondence to: ]{ xiaomu1982@163.com}
\affiliation{Hebei Provincial Key Laboratory of Photoelectric Control on Surface and Interface, School of Science, Hebei University of Science and Technology, Shijiazhuang, 050018, China}

\author{Xing-Tao An}
\email[Correspondence to: ]{ anxt2005@163.com}
\affiliation{Hebei Provincial Key Laboratory of Photoelectric Control on Surface and Interface, School of Science, Hebei University of Science and Technology, Shijiazhuang, 050018, China}
\affiliation{College of Physics, Hebei Normal University, Shijiazhuang 050024, China}
\affiliation{Key Laboratory for Microstructural Material Physics of Hebei Province, School of Science, Yanshan University, Qinhuangdao 066004, China}

\date{\today}

\begin{abstract}
It has been demonstrated that topological quantum spin Hall (QSH) state exist in twisted bilayers of transition metal dichalcogenides. However, a comprehensive theoretical characterization of the topological edge states remains a topic of interest and an unresolved issue. Here, the topological transport properties of the twisted WSe$_2$ bilayers are investigated. Beyond the conventional single QSH, we identify emergent double and quartuple quantum spin Hall states, hosting two and four pairs of counter-propagating helical edge channels respectively. Furthermore, the charge carriers in these edge states are not localized at edge but rather the high potential point of the moiré superlattice boundary, undergoing interlayer transitions and propagating forward continuously. We term these edge states as moiré edge states. These edge states can survive in non-magnetic disorder, with the robustness of double QSH states surpassing that of single QSH states. At a twisting angle of $2.45^\circ$, the transition between the single and double QSH states can be achieved by adjusting the gate on the surface. Based on this, we propose a five-terminal device to as a topological current devider. Our findings provide support for the development of dissipationless spintronics.

\end{abstract}

\maketitle

\sloppy

\textit{Introduction.} Since the discovery of the quantum Hall effect (QHE) \cite{klitzing1980new,thouless1982quantized}, significant attention has also been devoted to the study and development of the quantum anomalous Hall (QAH) effect and the quantum spin Hall (QSH) effect \cite{haldane1988model,chang2013experimental,kane2005quantum,kane2005z,bernevig2006quantum,bernevig2006quantum2,konig2007quantum,liu2008quantum,hasan2010colloquium}. The topological edge states of these topological phases exhibit excellent robust ballistic transport properties \cite{sheng2005nondissipative,sheng2006quantum,jiang2009numerical,an2013quantum,knez2014observation,du2015robust}, holding potential applications in low-power electronic devices and spintronics. Recent studies have revealed the presence of topological states in twisted bilayer WSe$_{2}$ and MoTe$_{2}$ \cite{wu2019topological,yu2020giant,zhai2020theory,devakul2021magic}. The periodic moiré superlattice formed by the twist can be theoretically nested using Bloch’s theorem. The theoretical prediction of the topological moiré band in this system has been proposed in 2019 \cite{wu2019topological}, and the reason for the topology of the miniband has been explained by the existence of topological layer-pseudospin Skyrmions. Furthermore, under the preserved spin-valley locking at the valence band edges of WSe$_{2}$ and MoTe$_{2}$ and time-reversal symmetry \cite{xiao2012coupled,liu2014erratum}, this system can ultimately exhibit the QSH effect \cite{He2025Topological}. Meanwhile, recent experiments have revealed the occurrence of fractional quantum anomalous Hall (FQAH) effect and fractional QSH effect in twisted bilayer MoTe$_{2}$ due to valley polarization driven by electron correlations \cite{park2023observation,cai2023signatures,reddy2023fractional,morales2024magic,wang2024fractional,zeng2023thermodynamic,xu2023observation,kang2024observation2}.

The QSH effect is characterized by Z$_{2}$ topological invariants, hosting helical edge states safeguarded by time-reversal symmetry \cite{kane2005quantum,kane2005z,fu2007topological,chiu2016classification,wen2017colloquium}. However, the question remains whether the number of edge states for each spin and the Hall conductance can exceed unity in the QSH effect. 
Recent experiments in twisted homobilayer WSe$_{2}$ have observed two pairs of helical edge states, exhibiting a quantized spin Hall conductance of 2 $e^2/h$ per edge \cite{kang2024observation}. Similarly, homobilayer MoTe$_{2}$ exhibits QSH states with 3 $e^2/h$ conductance per edge, corresponding to three pairs of helical edge states \cite{kang2024observation2}. These topological phases hosting multiple helical edge pairs are termed multiple QSH states. Such multiple QSH states cannot be described by the conventional Z$_{2}$ classification. Instead, high spin Chern numbers and emergent spin $U(1)$ quasisymmetry serve as essential factors sustaining quantized spin Hall conductance \cite{Lu2024Quantum, ezawa2013high,bai2022doubled,xue2023antiferromagnetic}. 
On the other hand, a topological current divider has been proposed as a key circuit element for the topological electronics and has been experimentally realized by using ferromagnetic electrode, disorder, or layer degree of freedom of magnetic topological insulator \cite{romeo2020topological,ovchinnikov2022topological,romeo2023experimental,zhang2021chiral,wu2021building}. This concept opens the door to the creation of topological programmable circuit board in which the domain wall currents can be split, rerouted, or switched off by tuning the Chern numbers of domains. However, the design of a topological current divider in QSH system with high spin Chern numbers (like twisted bilayer WSe$_2$ and MoTe$_2$) through weak gate voltages rather than intense magnetic fields or uncontrollable disorder remains a challenging work. This requires a systematic study of the topology and edge state distribution of the high spin chern number QSH systems.

In this Letter, we verified the existence of the multiple QSH state in twisted bilayer WSe$_{2}$ using Green’s function method, based on the Hamiltonian model of layer-pseudospin Skyrmions lattice \cite{wu2019topological}. Bulk band calculations reveal valley Chern number sequences $[1, -1, 2, 2]$ at $\theta = 1.16^\circ$ and $[1, 1, 1, 1]$ at $\theta = 2.45^\circ$, signifying an emergent spin Chern insulator with high spin Chern numbers. Subsequent nanoribbon simulations demonstrate minigap-localized topological edge states consistent with the spin Chern numbers, exhibiting moiré-confinement of carriers and interlayer-alternating current paths. Disorder robustness analysis confirms the multiple QSH state not only survives under weak disorder but also exhibits enhanced stability compared to the single QSH phase. 
By actively tuning the chemical potential, we achieve reversible switching between double and single QSH states in a heterostructure with $\theta = 2.45^\circ$ twist angle. This capability provides the fundamental technical basis for designing topological current splitter using surface gate architectures. Numerical results further verified that the boundary topological current remains robust even when traversing topological domain walls.

\begin{figure}
	\centering
	\includegraphics[width=9cm]{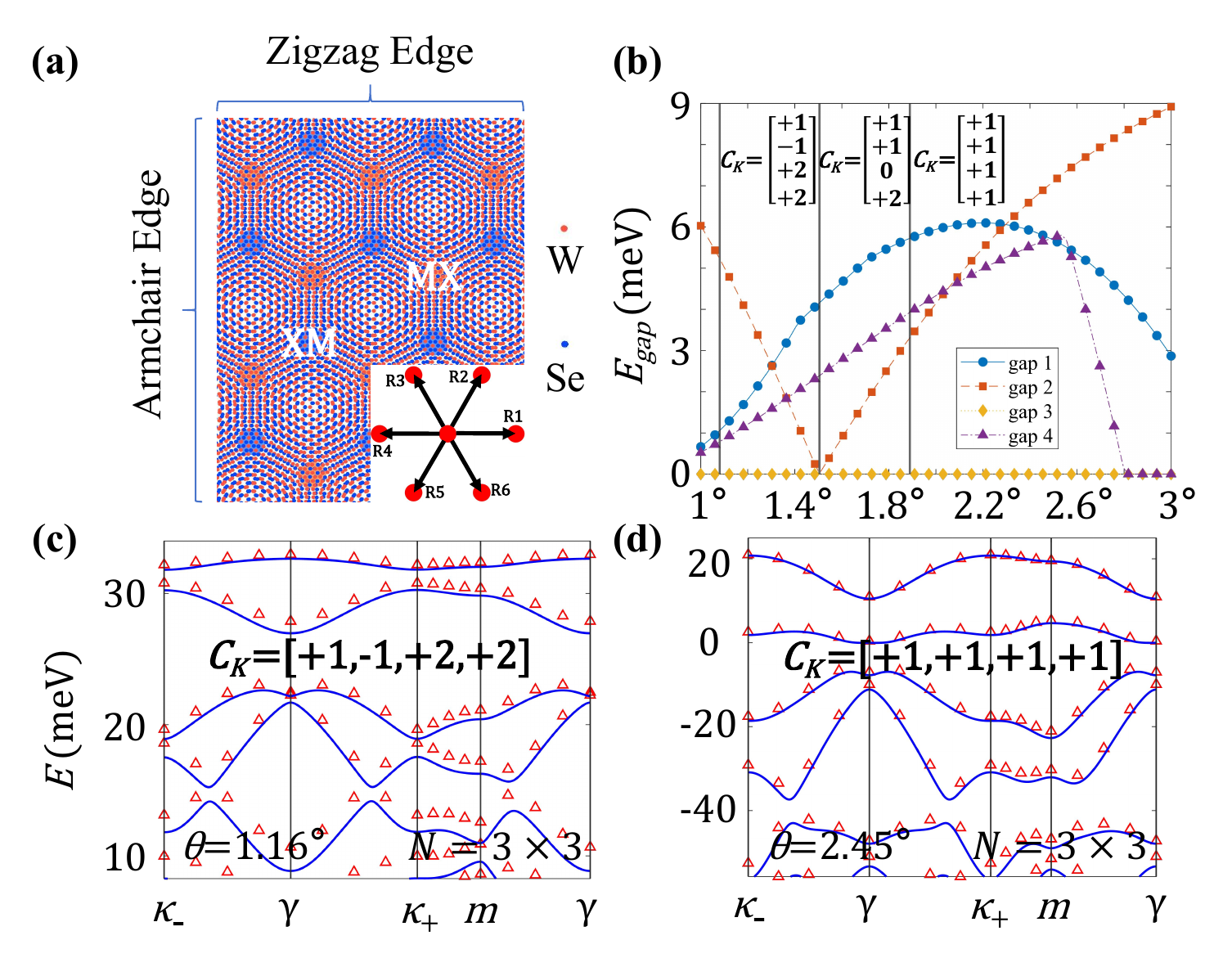}
	\caption{\label{fig:a1} (a) Schematic illustration of the moiré superlattice in twisted bilayer WSe$_2$, where red atoms represent W and blue atoms represent Se, forming a moiré superlattice due to the twist. The pink and light blue regions denote the moiré potential high points of the top and bottom layers respectively. Red dots in the white-background inset schematic denote lattice sites of the TB model, with nearest-neighbor vector directions aligned to the moiré superlattice orientations depicted in the main figure. (b) Energy gaps of moiré minibands as a function of twist angle. Gap indices 1–4 represent the band gaps between the first to fifth minibands. Valley Chern numbers of the four highest minibands are annotated at respective twist angles. (c) Calculated moiré band structure at $\theta = 1.16^\circ$ and (d) $\theta = 2.45^\circ$. Blue curves denote continuum-model calculations, while red triangles represent TB computations employing parameterization $N = 3 \times 3$.
	}
\end{figure}
\textit{Model and methods.} Figures.~\ref{fig:a1}(a) schematically illustrates the moiré structure of twisted bilayer WSe$_2$. Here, high-symmetry stacking regions MX and XM are denoted by red and blue markers, respectively (where M represents metal atoms and X denotes chalcogen atoms, with atomic columns vertically aligned in these high-symmetry regions). We treat the MX and XM stacking regions as lattice points, whose interconnections form an approximate hexagonal lattice. The Hamiltonian of a homogeneous twisted bilayer TMDs system can be described using a continuum model as \cite{wu2019topological}
\begin{equation}\label{eqi1}
	H_{K} = \begin{pmatrix}
		-\frac{\hbar^2 (\mathbf{k} - \mathbf{\kappa}_{+})^2}{2m^*} + V_1(\mathbf{r} ) & V_T(\mathbf{r} ) \\
		V_T^{\dagger}(\mathbf{r} ) & -\frac{\hbar^2 (\mathbf{k} - \mathbf{\kappa}_{-})^2}{2m^*} + V_2(\mathbf{r} )
	\end{pmatrix}.
\end{equation}
The Hamiltonian characterizes the spin-up component of the $K$ valley. This representation leverages time-reversal symmetry connecting the $K$ and $-K$ valleys, where strong spin-orbit coupling (SOC) and monolayer WSe$_2$'s inversion symmetry breaking collectively enforce spin-valley locking assigning defined spin orientations to individual valleys in the twisted bilayer system.
$H_{K}$ comprises kinetic and potential terms for the top and bottom layers, along with the interlayer tunneling term. Due to the presence of the twist, there is a momentum space mismatch between the $K$ valleys of the top and bottom layers, leading to the emergence of $\kappa_-=(\mathbf{g}_1+\mathbf{g}_6)/3$ and $\kappa_+=(\mathbf{g}_1+\mathbf{g}_2)/3$. Here $\mathbf{g}_i = \frac{4\pi}{\sqrt{3} a_M}(\cos(\frac{\pi(i-1)}{3}),
\sin(\frac{\pi(i-1)}{3}))$ represents the moiré reciprocal lattice vectors, where $a_M$ is the moiré period that is closely related to the twisting angle and satisfies $a_M = \frac{a_0}{2 \sin\left(\frac{\theta}{2}\right)} \approx \frac{a_0}{\theta}$ ($a_0$ represents the lattice constant of monolayer WSe$_2$). $V_1(\mathbf{r})$, $V_2(\mathbf{r})$ and $V_T(\mathbf{r})$ represent the moiré potentials tern and interlayer tunneling term, respectively
\begin{equation}\label{eqi2}
	V_l(\mathbf{r}) = -2V \sum_{i=1,3,5} \cos(\mathbf{g}_i \cdot \mathbf{r} + \phi_l), 
\end{equation}
\begin{equation}\label{eqi3}
	V_T(\mathbf{r}) = \omega(1+e^{i\mathbf{g}_2 \cdot \mathbf{r}}+e^{i\mathbf{g}_3 \cdot \mathbf{r}}).
\end{equation}
where $l=1$ for the top and $l=2$ for the bottom layers. The $V$ and $\omega$, as parameters, respectively determine the strength of the intralayer moiré potential and the intensity of interlayer tunneling. Additionally, $\phi_l$ determines the shape of the intralayer moiré potential. The relevant parameters were initially determined by local stacking calculations \cite{wu2019topological}, and subsequently refined by large-scale first-principles calculations to obtain new and more accurate parameters \cite{devakul2021magic,reddy2023fractional,wang2024fractional}. In this Letter, the parameter values we use are: $\phi_1=-\phi_2=128^{\circ} $, $V=9meV$, $\omega=-18meV$, $m=0.43m_e$, $a_0=3.32\text{\AA} $ \cite{devakul2021magic,reddy2023fractional}. In this Hamiltonian framework, we have neglected the possible intervalley scattering, which typically occurs in the conduction band edge \cite{an2017realization} and leads to the inability of edge states to survive in weak disorder \cite{khalifa2023absence}. Moreover, due to the spin-valley locking in twisted bilayer WSe$_2$\cite{xiao2012coupled}, intervalley scattering is almost prohibited in the valence band edge, therefore it can be safely neglected.

To analyze the transport details of edge states, we employ the Green's function method for calculations. However, a continuum model is inherently incompatible with Green's function computations, necessitating the transformation of the continuum model into a tight-binding (TB) model. To preserve the $C_{3z}$ symmetry of the system, we implement the TB model on a triangular lattice. Utilizing the finite-difference framework, the TB model constructed on this triangular lattice can be formulated as
\begin{equation}\label{eqi4}
	\begin{split}
		H_{b-TB}=
		\sum_{i,l}\left[\varepsilon_l+V_l\left(\mathbf{r}_i\right)\right]c_{i,l}^\dagger\mathrm{c}_{i,l} +\sum_{i}V_T\left(\mathbf{r}_i\right)c_{i,2}^\dagger& c_{i,1} \\ 
		+\sum_{i}V_T^\dagger\left(\mathbf{r}_i\right)c_{i,1}^\dagger c_{i,2}+\sum_{<i,j>,l}t_{i,j,l}c_{i,l}^\dagger c_{j,l},
	\end{split}
\end{equation}
	\noindent where $\varepsilon_l$ and $V_l(\mathbf{r}_i)$ is the on-site energy and moiré potential at $\mathbf{r}_i$,  $c_{i,l}^\dagger$ is the creation operator for electron at sit $i$ on layer $l$. $t_{i,j,l}$ is the hopping between sites $i$ and $j$ on layer $l$, and $\braket{,}$ denotes summation over nearest-neighbor pairs only. In the triangular lattice, there are six nearest-neighbor around one site, and the hopping matrix for each nearest neighbor can be expressed as
	 \begin{equation}\label{eqi5}
	 	\begin{aligned}
	 		t_1(\mathbf{R1})=
	 		\begin{bmatrix}
	 			te^{-i\frac{2\pi}{3N}} & 0 \\
	 			0 & te^{-i\frac{4\pi}{3N}} 
	 		\end{bmatrix}, 
	 	\end{aligned}
	 \end{equation}
     \begin{equation}\label{eqi5}
     	\begin{aligned}
     		t_2(\mathbf{R2})=
     		\begin{bmatrix}
     			te^{-i\frac{4\pi}{3N}} & 0 \\
     			0 & te^{-i\frac{2\pi}{3N}} 
     		\end{bmatrix}, 
     	\end{aligned}
     \end{equation}
     \begin{equation}\label{eqi5}
     	\begin{aligned}
     		t_3(\mathbf{R3})=
     		\begin{bmatrix}
     			te^{-i\frac{2\pi}{3N}} & 0 \\
     			0 & te^{i\frac{2\pi}{3N}} 
     		\end{bmatrix}, 
     	\end{aligned}
     \end{equation}
     $t_4(\mathbf{R4}) = t_1(\mathbf{R1})^\dagger$, $t_5(\mathbf{R5}) = t_2(\mathbf{R2})^\dagger$, and $t_6(\mathbf{R6}) = t_3(\mathbf{R3})^\dagger$, where $\mathbf{R1}$ to $\mathbf{R6}$ are the six vectors connecting the nearest lattice sites in the TB model, as shown in Fig.~\ref{fig:a1}(a). Here, $t = \frac{\hbar^2}{3m^*a}$, with $a$ being the lattice constant of the TB model, defined as the distance between adjacent sites. The lattice constant $a$ can be determined by $a = a_M / N$, where $N$ is the number of grid points along the primitive lattice vectors of the moiré unit cell. To ensure that all high-symmetry points within a moiré unit cell are included, $N$ is typically chosen as a multiple of $3$. Finally, the on-site energy is given by $\varepsilon_l = -6t$.

\begin{figure}
	\centering
	\includegraphics[width=8.5cm]{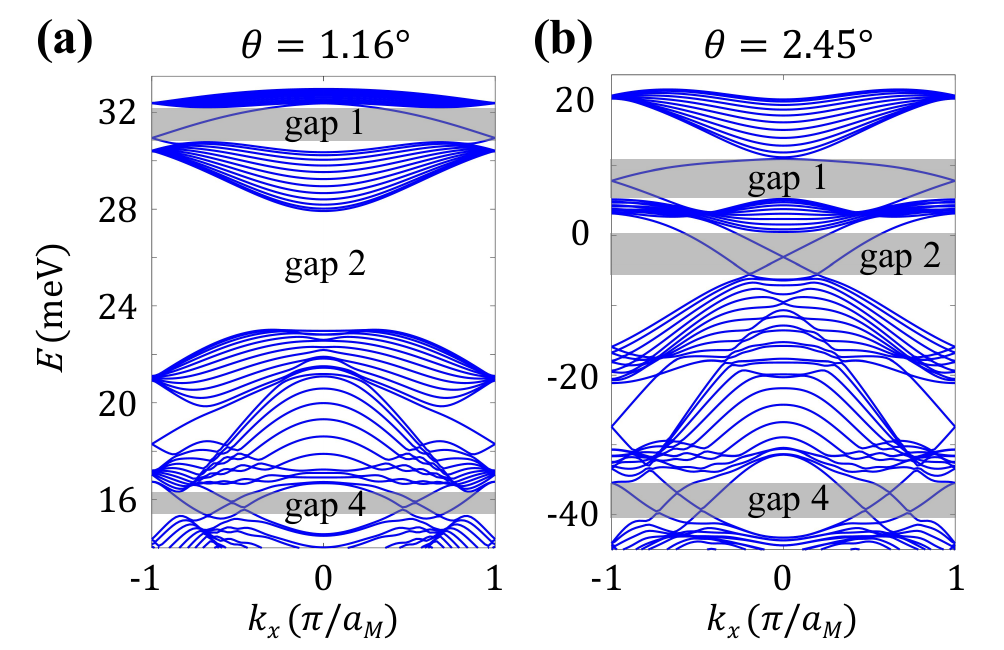}
	\caption{\label{fig:a2}(a) Zigzag-edged nanoribbon band structure at twist angle $\theta = 1.16^\circ$, and (b) at $\theta = 2.45^\circ$. Shaded areas indicate energy regions hosting topological edge states within the bulk gap.
	}
\end{figure}

\begin{figure*}
	\centering
	\includegraphics[width=15.5cm]{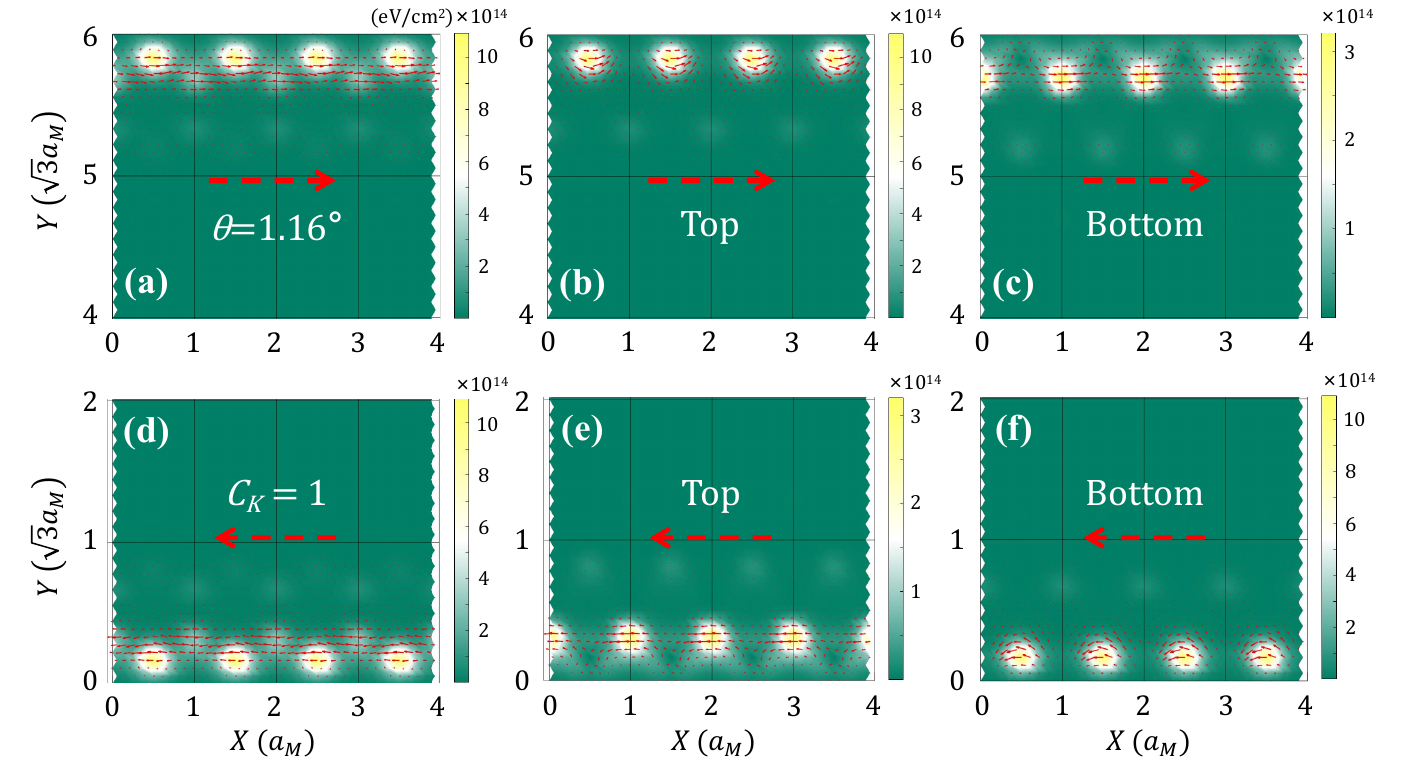}
	\caption{\label{fig:a3} LDOS and LCD distributions for topological edge states within gap 1 at $\theta = 1.16^\circ$ (Valley Chern number $C_K=1$ phase). (a) and (d) Full-system LDOS and LCD; (b) and (e) Layer-resolved LDOS and LCD in the top layer; (c) and (f) Corresponding distributions in the bottom layer. Red arrows indicate unidirectional carrier transport along chiral edge channels.
	}
\end{figure*}

\begin{figure*}
	\centering
	\includegraphics[width=15.5cm]{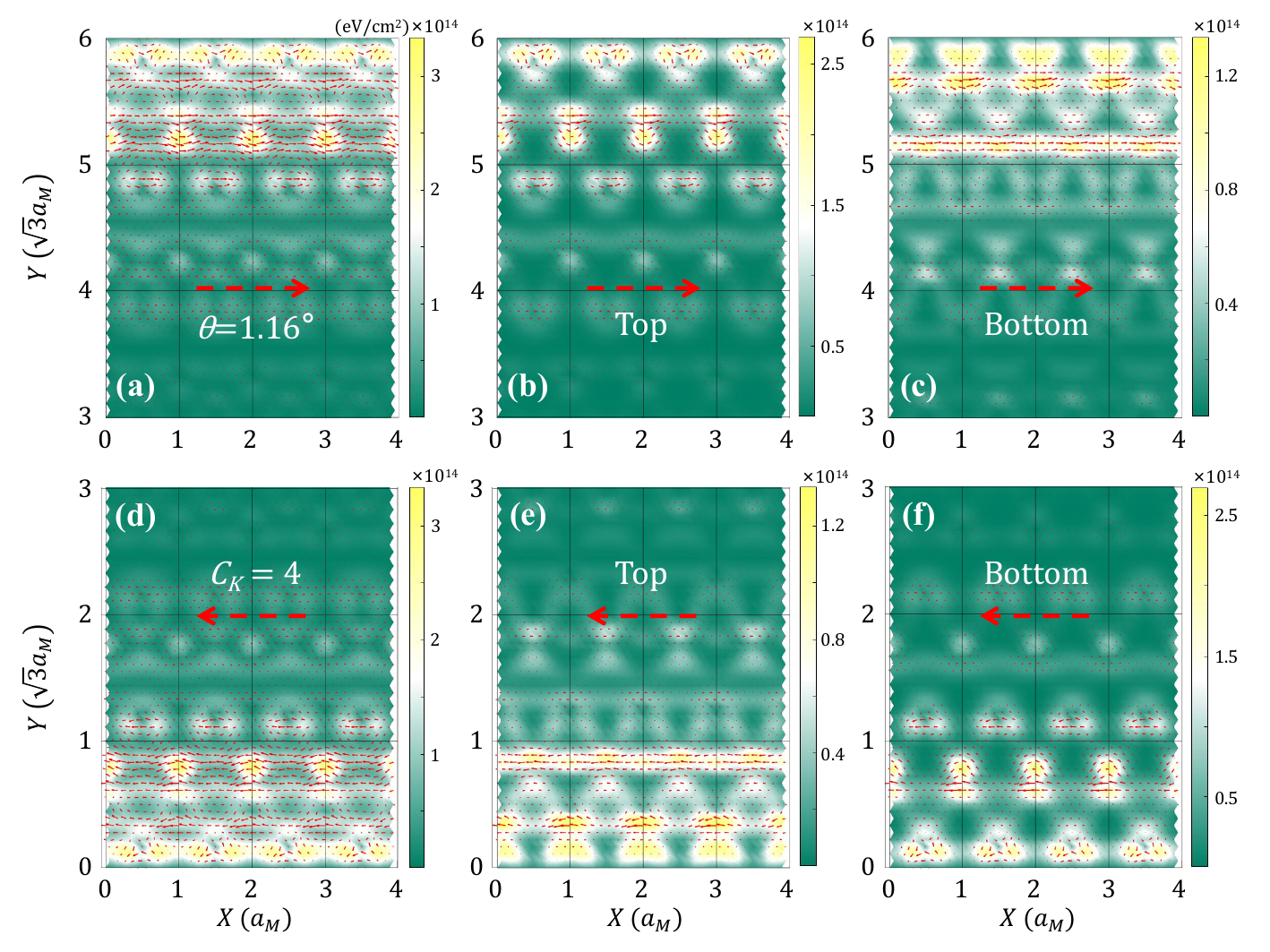}
	\caption{\label{fig:a4} LDOS and LCD distributions for topological edge states within gap 4 at $\theta = 1.16^\circ$ (Valley Chern number $C_K=4$ phase). (a) and (d) Full-system LDOS and LCD; (b) and (e) Layer-resolved LDOS and LCD in the top layer; (c) and (f) Corresponding distributions in the bottom layer. Red arrows indicate unidirectional carrier transport along chiral edge channels.
	}
\end{figure*}

\begin{figure*}
	\centering
	\includegraphics[width=15.5cm]{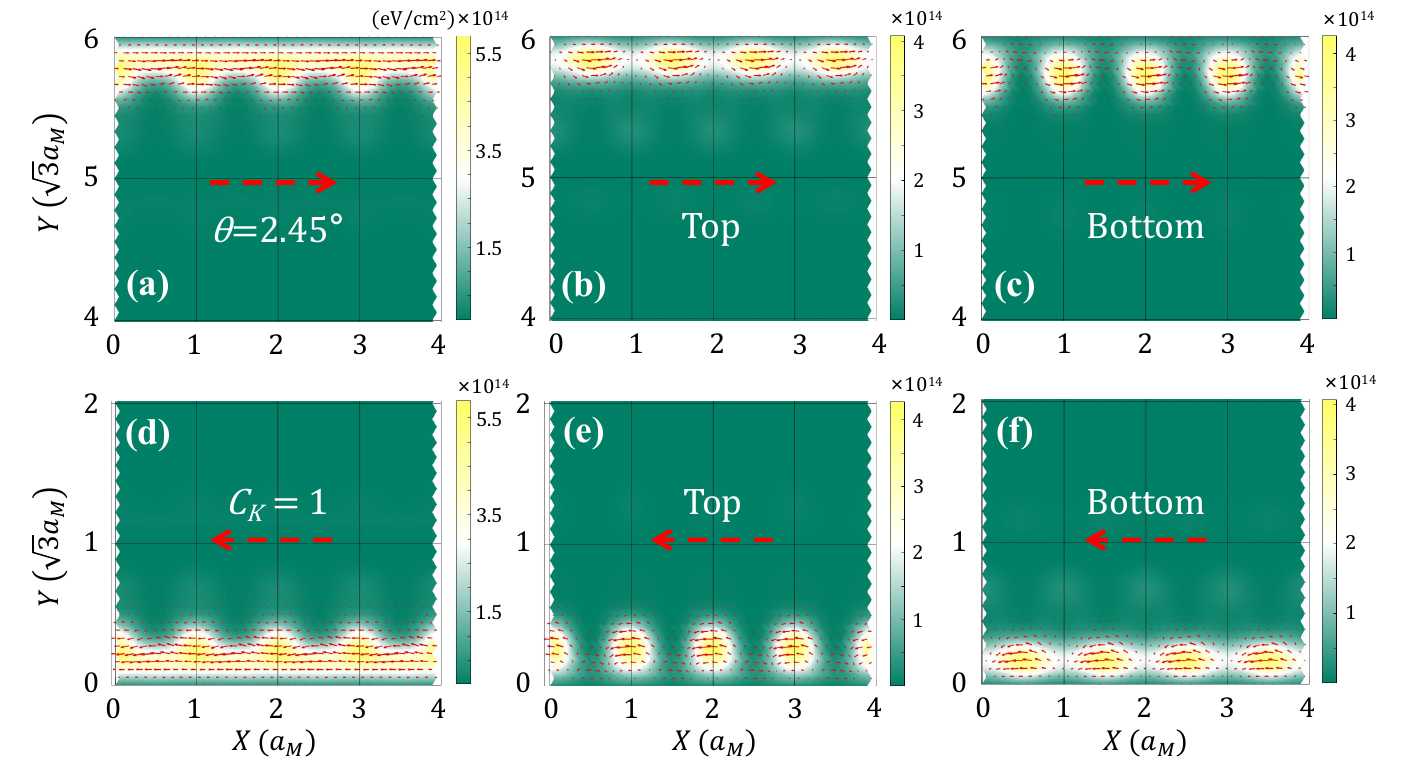}
	\caption{\label{fig:a5} LDOS and LCD distributions for topological edge states within gap 1 at $\theta = 2.45^\circ$ (Valley Chern number $C_K=1$ phase). (a) and (d) Full-system LDOS and LCD; (b) and (e) Layer-resolved LDOS and LCD in the top layer; (c) and (f) Corresponding distributions in the bottom layer. Red arrows indicate unidirectional carrier transport along chiral edge channels.
	}
\end{figure*}

\begin{figure*}
	\centering
	\includegraphics[width=15.5cm]{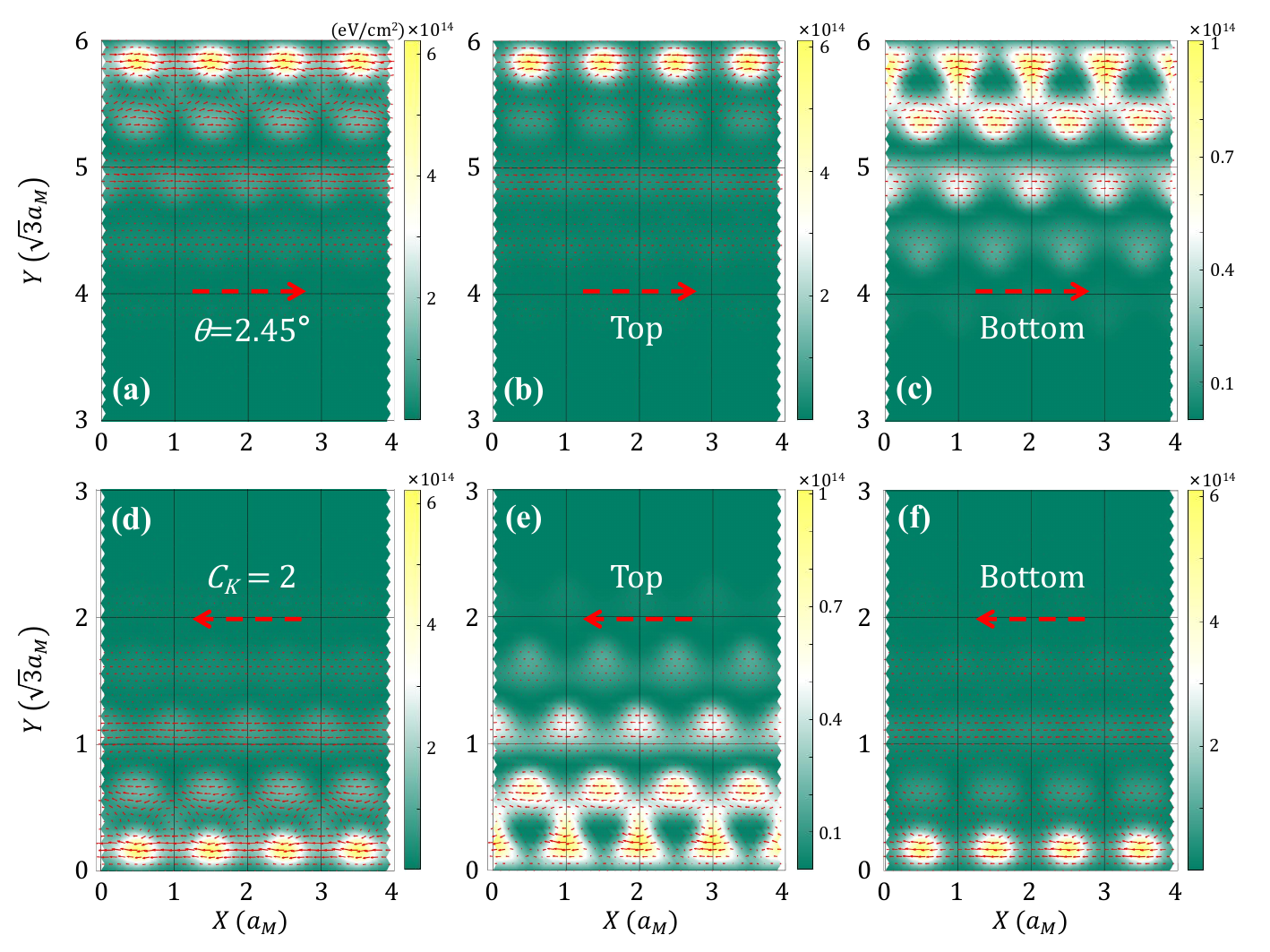}
	\caption{\label{fig:a6} LDOS and LCD distributions for topological edge states within gap 2 at $\theta = 2.45^\circ$ (Valley Chern number $C_K=2$ phase). (a) and (d) Full-system LDOS and LCD; (b) and (e) Layer-resolved LDOS and LCD in the top layer; (c) and (f) Corresponding distributions in the bottom layer. Red arrows indicate unidirectional carrier transport along chiral edge channels.
	}
\end{figure*}

We employ the Green's function approach to study the transport properties of the twisted bilayer WSe$_2$. We consider a zigzag moiré superlattice nanoribbon, as shown in the Fig.~\ref{fig:a1}(a), where the left source lead and right drain lead are both semi-infinitely long superlattice nanoribbons, and the central scattering region with a length of ${L}$ and width of ${W}$ is the focus of our study. In this case, the lateral minimum period of the nanoribbon is $a_M$. The transmission probability of electrons with energy $E$ is given as \cite{fisher1981relation}
\begin{equation}\label{eqi4}
	T(E) = \text{Tr}[\Gamma_L(E)G(E)\Gamma_R(E)G^\dagger(E)].
\end{equation}
The Green's function $G(E)$ satisfies the equation
\begin{equation}\label{eqi5}
	(EI-H-\Sigma_L(E)-\Sigma_R(E))G(E) = I, 
\end{equation}
where, $H$ is the Hamiltonian in the scattering region, and $I$ denotes the identity matrix. The left self-energy $\Sigma_L(E)$ and right self-energy $\Sigma_R(E)$ can be derived from the surface Green's functions of the two terminals calculated using a rapid convergence method \cite{sancho1985highly}. The linewidth functions can be obtained from the self-energy,
\begin{equation}\label{eqi6}
    \Gamma_L (E)=i(\Sigma_L(E)-{\Sigma_L(E)}^\dagger),
\end{equation}
\begin{equation}\label{eqi7}
	\Gamma_R (E)=i(\Sigma_R(E)-{\Sigma_R(E)}^\dagger).
\end{equation}
In the framework of the quasi-equilibrium Green's function, the conductance is given by 
\begin{equation}\label{eqi8}
	G(E) = \frac{e^2}{h} T(E).
\end{equation}
The local density of states (LDOS) and local current density (LCD) at $i$ point for transport from left to right can be expressed as \cite{meir1992landauer}
\begin{equation}\label{eqi9}
	n_i (E)=[(G(E) \Gamma_L (E)G(E)^\dagger)/\pi]_{i,i}
\end{equation}
and
\begin{align}\label{eqi10}
	J_{i,i+1} = & \frac{2e}{h} \left\{ H_{i,i+1} \text{Im}[G(E)\Gamma_L(E)G(E)^\dagger]_{i,i+1} \right. \nonumber \\
	& -H_{i+1,i}\left.\text{Im}[G(E)\Gamma_L(E)G(E)^\dagger]_{i+1,i} \right\}.
\end{align} 
In the case of right to left transport, replace $\Gamma_L(E)$ with $\Gamma_R(E)$ in the expression above.

\begin{figure*}
	\centering
	\includegraphics[width=15.7cm]{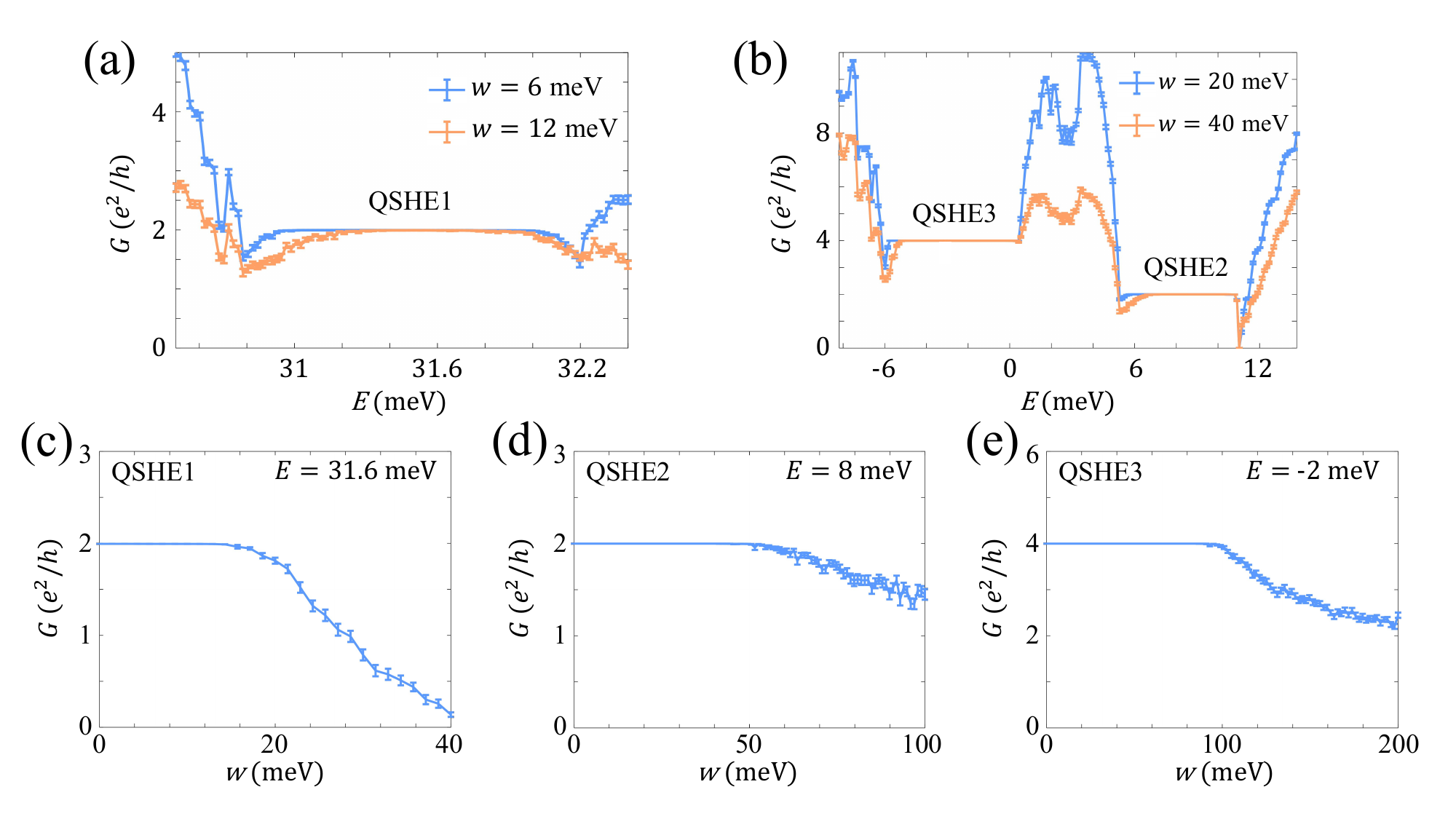}
	\caption{\label{fig:a7}(a) The relationship between conductance and electron incident energy at $\theta = 1.16^\circ$ and disorder strengths of $w=6$ meV and $w=12$ meV. QSHE1 resides within gap 1. (b) The relationship between conductance and electron incident energy at $\theta = 2.45^\circ$ and disorder strengths of $w=20$ meV and $w=40$ meV. QSHE2 resides within gap 1, while QSHE3 occupies gap 2. (c), (d), and (e) investigate the functional relationship between conductance and disorder strength for electron incident energies set at the positions "QSHE1", "QSHE2", and "QSHE3", respectively.
	}
\end{figure*}

\textit{Topological moiré bands.} Based on both continuum and TB models, we calculate the band structures at the K valley for two small twist angles. Figures.~\ref{fig:a1}(c) and~\ref{fig:a1}(d) display band structures for commensurate angles $\theta = 1.16^\circ$ and $\theta = 2.45^\circ$, where blue curves and red triangles denote results from the continuum and TB models with grid density $N = 3 \times 3$, respectively. Excellent agreement is observed between the two models, and further increasing $N$ enhances the accuracy of TB model. We maintain $N = 3 \times 3$ in subsequent calculations to reduce computational cost. Valley Chern numbers for the first four moiré bands are calculated as $C_K=[1,-1,2,2]$ at $\theta = 1.16^\circ$ and $C_K=[1,1,1,1]$ at $\theta = 2.45^\circ$ (Related via time-reversal symmetry, the valley Chern numbers satisfy $C_K=-C_{-K}$), with identical values across models. This confirms the TB model accurately captures all low-energy properties. The two twist angles are representatives of the twisted bilayer WSe$_2$ possessing topological phases in the small-angle regime. Supporting evidence appears in Fig.~\ref{fig:a1}(b), where we plot inter-band gaps (gap 1–gap 4) versus twist angle. Since bands 3 and 4 remain gapless at all twist angles, we need only consider changes in the sum of Chern numbers (bands 3 and 4). Consequently, the phase characterized by the Chern vector $C_K=[1,1,0,2]$ is topologically equivalent to the phase with $C_K=[1,1,1,1]$. A gap close and reopen between bands 2 and 3 indicates a topological phase transition, with corresponding valley Chern numbers annotated for each phase (see Fig~\ref{fig:a1}(c)and~\ref{fig:a1}(d)).

\textit{Multiple quantum spin Hall states.} The emergence of a high spin Chern number signifies the formation of multiple QSH states in the system. To investigate electronic transport in twisted bilayer WSe$_2$, we construct moiré-aligned zigzag-edged nanoribbons using the TB model and calculate their band structures for width $W=6\sqrt{3}a_M$ at twist angles $\theta = 1.16^\circ$  (Fig.~\ref{fig:a2}(a)) and $\theta = 2.45^\circ$ (Fig.~\ref{fig:a2}(b)), where bulk-edge correspondence reveals quantum spin Hall edge states across multiple gaps: at $\theta = 1.16^\circ$, one pair of helical edge states emerges between the first and second moiré bands (total Chern number $C_K = 1$ above gap 1), no edge states exist between bands 2 and 3 ($C_K = 0$ above gap 2), and four pairs of helical edge states develop between bands 4 and 5 ($C_K = 4$ above gap 4); at $\theta = 2.45^\circ$, two pairs of helical edge states form between bands 2 and 3 ($C_K = 2$ above gap 2) due to topological phase transition.
\begin{figure*}
	\centering
	\includegraphics[width=18cm]{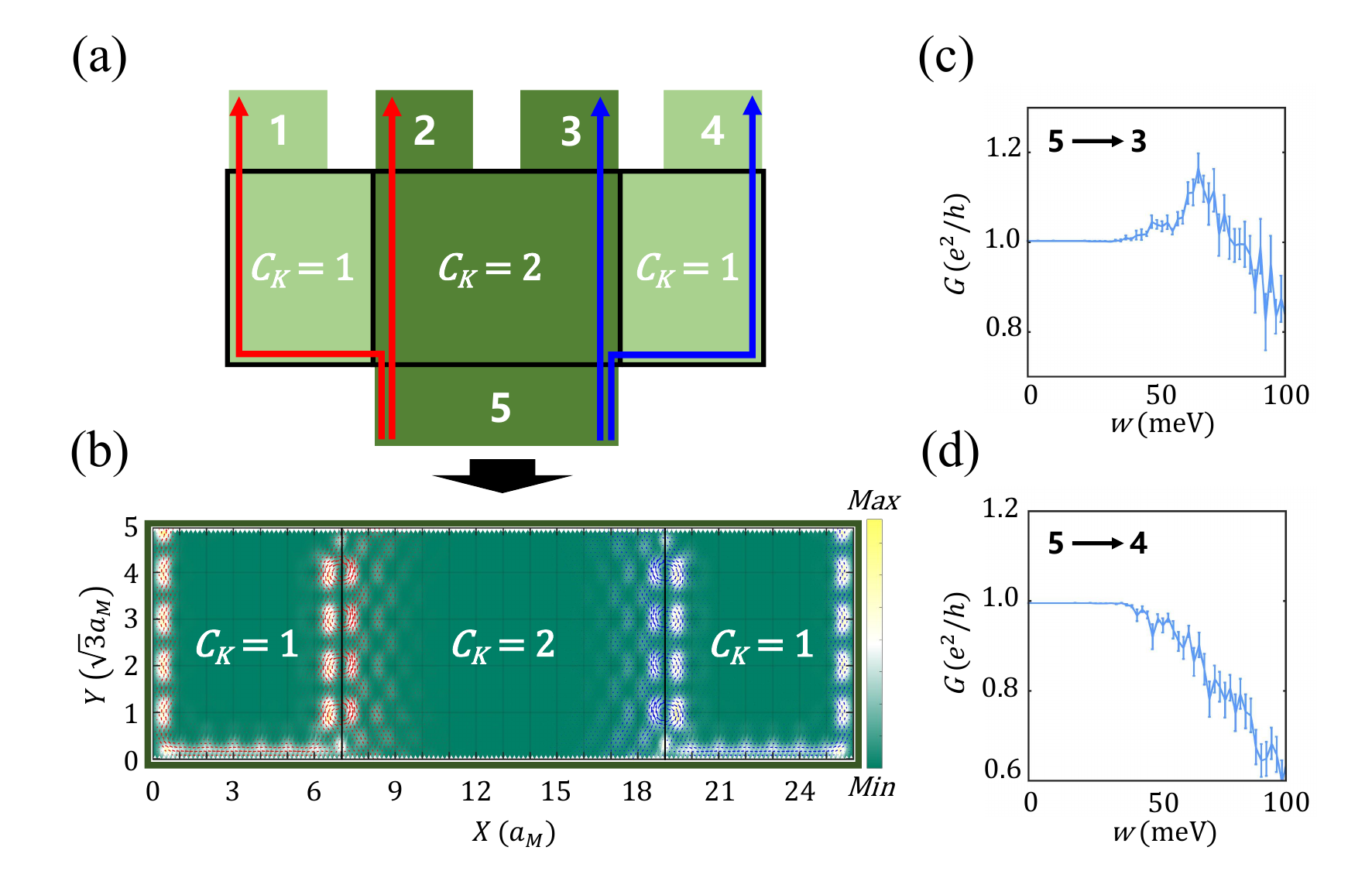}
	\caption{\label{fig:a8}(a) Schematic diagram of a five-terminal topological current divider. The red arrows represent the flow of spin-up currents, while the blue arrows indicate the flow of spin-down currents. (b) Depicts the calculated LDOS and LCD within the central scattering region of the device. The widths of the three domains are $7a_M$, $10a_M$, and $7a_M$ from left to right, respectively. (c) and (d) depict the terminal 5 to terminals 3 and 4, respectively, as a function of disorder strength.
	}
\end{figure*} 

Based on edge states in the band structure, we employ the Green's function method to compute the LDOS and LCD for edge states within various gaps. Considering time-reversal symmetry and spin-valley locking of the two valleys of twisted bilayer WSe$_2$, we calculate only one spin channel. While spin-up carriers propagate from left to right along the nanoribbon's upper edge (see Fig.~\ref{fig:a3}(a)), their time-reversal counterparts spin-down carriers traverse from left to right along the lower edge, constituting counter-propagating helical edge channels. As shown in Fig.~\ref{fig:a3} at $\theta = 1.16^\circ$ with incident energy in gap 1 ($C_K = 1$ phase), we present both layer-resolved LDOS and LCD. Carriers persistently localize at high-symmetry points of maximal moiré potential in each layer (MX and XM regions) — a direct consequence of enhanced the approximately flat bands near gap1 at small twist angles that intensify moiré-induced localization. 
Carriers propagate not along geometric edges but rather follow trajectories tracing moiré potential maxima, establishing a distinct moiré edge state. Layer-resolved analyses in Fig.~\ref{fig:a3}(b),(c) and~\ref{fig:a5}(e),(f) elucidate the underlying transport mechanism: Continuous interlayer tunneling facilitates carrier transport — when carriers accumulate at high-potential sites in the top layer (MX regions), they quantum-mechanically tunnel to adjacent high-potential sites in the bottom layer (XM regions) through the $V_T$ term. This process generates spatially alternating interlayer current pathways.
Given that carriers in the top moiré bands exhibit honeycomb-like localization patterns with valley Chern numbers $C_K = [1,-1]$, the helical edge states can often be reliably modeled using a two-band Kane-Mele Hamiltonian, yielding quantum spin Hall conductance of 2 $e^2/h$ at filling $v=2$, consistent with experimental observations.
At $\theta = 1.16^\circ$ with energy in gap 4 ($C_K = 4$ phase, Fig.~\ref{fig:a4}), the system exhibits a conductance of 8 $e^2/h$ (Here we consider the sum of contributions from both spins.). 
In contrast to the strong localization observed in gap 1, layer-resolved data presented in Fig.~\ref{fig:a3}(b),(c) and~\ref{fig:a5}(e),(f) demonstrate a significant delocalization of carrier spatial distributions. The LDOS no longer strictly confines carriers to high-symmetry stacking sites of the moiré potential. This phenomenon arises from increased band curvature in deeper moiré bands (Bands 3-4 in Fig.~\ref{fig:a1}(c)), which attenuates the quantum confinement effects induced by the moiré potential.
Significantly, LCD measurements reveal intricate current pathways originating from phase-coherent superposition of wavefunctions belonging to four helical edge state pairs at identical energy but distinct momenta. This superposition generating a networked transport topology characterized by spatially braided current flows, as directly visualized in Fig.~\ref{fig:a5} (a) and~\ref{fig:a5}(d).
Fig.~\ref{fig:a5} and Fig.~\ref{fig:a6} show results at $\theta = 2.45^\circ$ for gap 1 ($C_K = 1$, 2 $e^2/h$) and gap 2 ($C_K = 2$, 4 $e^2/h$), respectively. Larger twist angles broaden carrier localization, invalidating the Kane-Mele approximation and necessitating higher-resolution lattice models.

Next, we studied robustness of the multiple QSH state by investigating the influence of nonmagnetic disorder on the conductance of the edge states, as depicted in Fig.~\ref{fig:a7}. We randomly sample energy from $-w/2$ to $w/2$ and add it to the on-site energy at each lattice site, defining disorder strength as $w$. All data were calculated thirty times and averaged, with errors reported. As shown in Fig.~\ref{fig:a7}(a), at the position of "QSHE1 (quantum spin Hall effect 1)" ($\theta = 1.16^\circ$, gap 1), the bulk states are scattered off by disorder at $w=6$ meV, while the conductance of the edge states remain quantized (2 $e^2/h$). When disorder increases to $w=12$ meV, the bulk states are more prominently scattered, while the edge states plateau remains stable. At this point, with the energy set at "QSHE1" and increasing disorder strength continuously, as shown in Fig.~\ref{fig:a7}(c), the conductance plateau remains stable over a significant range of $w$ values. Scattering sets in near $w \approx 14$ meV, beyond which the conductance persistently diminishes with further disorder enhancement. It is worth noting that the minigap at "QSHE1" is only about $1.4$ meV, and the magnitude of the moiré potential is approximately around 30 meV. Nevertheless, this QSH state can survive under disorder with a maximum strength of $w \approx 14$ meV. As shown in Fig.~\ref{fig:a7}(b), "QSHE2"($\theta = 2.45^\circ$, gap 1) and "QSHE3"($\theta = 2.45^\circ$, gap 2) also exhibit stable conductance plateaus (2 $e^2/h$ and 4 $e^2/h$ ) at $w = 20$ meV and $w = 40$ meV disorder strengths. 
Notably, at a twist angle of $\theta = 2.45^\circ$, the single QSH state exhibits a minigap of approximately $5.7$ meV at "QSHE2", maintaining a stable conductance plateau up to disorder strengths of $w \approx 50$ meV. In contrast, the double QSH state at "QSHE3" features a approximately $6.1$ meV minigap, with its conductance plateau remaining robust against disorder up to $w \approx 100$ meV — twice the stability range of the single QSH phase, as demonstrated in Figs.~\ref{fig:a7}(d) and~\ref{fig:a7}(e). 
The robustness of the QSHE at different twist angles is fundamentally determined by the miniband gap width, wherein larger gaps exhibit greater stability, requiring stronger disorder intensity to generate in-gap states within the bulk spectrum.

\textit{Gate-controlled spin-resolved topological current divider.} Helical edge states with varying multiplicities localized within distinct minigaps enable gate-tunable transitions between different QSH phases. This capability facilitates the design of versatile topological devices. Here, we design a five-terminal device based on the multiple QSH states of twisted bilayer WSe$_2$, capable of achieving topological current splitting and spin current splitting, as illustrated in Fig.~\ref{fig:a8}(a). Under a twist angle of $\theta = 2.45^\circ$, gate voltage modulation elevates the Fermi level to gap 1 in the central region and terminals 2, 3, 5, while maintaining the Fermi level in gap 2 for side regions and terminals 1, 4. This configuration establishes the central region in the $C_K = 2$ phase and side regions in the $C_K = 1$ phase. Thus, spin-up current propagates along the left boundary while spin-down current transports along the right boundary. Crucially, one edge state of the $C_K = 2$ phase spatially overlaps with the corresponding edge state of the $C_K = 1$ phase. Given that the $C_K = 2$ phase hosts two edge channels and the $C_K = 1$ phase has only one, half of the current from the $C_K = 2$ phase diverges into the $C_K = 1$ phase, enabling topologically protected current splitting across the domain wall. The computational results in Fig.~\ref{fig:a8}(b) validate this mechanism’s feasibility. Notably, current flowing from terminal 5 to terminals 2 and 3 cyclically traverses the topological boundary, generating a propagating vortex within the counter-propagating edge channels. Figures.~\ref{fig:a8}(c) and~\ref{fig:a8}(d) demonstrate the robustness of this topological splitting against disorder. At disorder strengths of approximately $w \approx 40$ meV, edge states maintain quantized conductance, confirming exceptional disorder tolerance. 
As disorder strength increases, the enhanced stability of the double QSH state causes carriers in terminal 4 to scatter more rapidly into bulk states and become confined within the double QSH phase. Consequently, the conductance of terminal 5 exhibits a transient enhancement before progressively declining at higher disorder levels. Ultimately, both terminals 4 and 5 exhibit conductances below their pristine-state values under strong disorder.

\textit{Conclusions.} In summary, we systematically investigate the QSH states in twisted bilayer WSe$_2$. Using a continuum model coupled with a finite-difference TB formalism, we calculate the moiré band structures at the $K$ valley for twist angles $\theta = 1.16^\circ$ and $\theta = 2.45^\circ$. Valley Chern number analysis reveals topological invariants $C_K=[1,-1,2,2]$ for the first four moiré bands at $\theta = 1.16^\circ$, while $C_K=[1,1,1,1]$ emerges at $\theta = 2.45^\circ$. Corresponding helical edge states—whose channel counts match the respective Chern numbers—are identified within the band gaps through nanoribbon band structure calculations. Green's function computations of LDOS and LCD distributions demonstrate moiré-edge-localized carrier confinement and interlayer current tunneling dynamics. These QSH phases exhibit robustness against non-magnetic disorder, with high spin Chern number states showing significantly enhanced disorder tolerance. Leveraging gate-tunable topological phase control at fixed twist angles, we design a five-terminal topological splitter capable of concurrent current branching and spin-resolved separation, whose functionality persists under disorder as confirmed by non-equilibrium quantum transport simulations. This work establishes an efficient single-particle theoretical framework for probing topological transport in moiré superlattices and provides new design paradigms for twistronic spintronic devices based on engineered moiré potentials.

\textit{Acknowledgments.} This Letter was supported by the Scientific Research Project of Universities in Hebei Province (No. JCZX2025023), the National Natural Science Foundation of China (No. 12074096), and the Innovation Leading Talent Team Project in Hebei Universities. J.L. acknowledge support by the National Natural Science Foundation of China (No. 12274305).

\bibliography{mymy}

\end{document}